\documentclass[a4paper]{jpconf}
\usepackage{graphicx}
\newcommand{\bc}{\begin{center}}
\newcommand{\ec}{\end{center}}
\newcommand{\bt}{\begin{tabular}}
\newcommand{\et}{\end{tabular}}
\newcommand{\be}{\begin{equation}}
\newcommand{\ee}{\end{equation}}
\newcommand{\bea}{\begin{eqnarray}}
\newcommand{\eea}{\end{eqnarray}}
\newcommand{\bfig}{\begin{figure}}
\newcommand{\efig}{\end{figure}}
\newcommand{\ba}{\begin{array}}
\newcommand{\ea}{\end{array}}

\newcommand{\bfr}{\begin{flushright}}
\newcommand{\efr}{\end{flushright}}
\newcommand{\bfl}{\begin{flushleft}}
\newcommand{\efl}{\end{flushleft}}
\begin{document}

\title{Fermion Masses from Six Dimensions and Implications for Majorana
Neutrinos}

\author{J-M Fr\`{e}re$^1$, M Libanov$^2$, S Mollet$^1$ and S Troitsky$^2$}

\address{$^1$ Physique Th\'{e}orique CP 225, Universit\'{e} Libre de Bruxelles, B-1050 Bruxelles, Belgium}
\address{$^2$ Institute for Nuclear Research of the Russian Academy of Sciences,
60th October Anniversary Prospect 7a, 117312, Moscow, Russia}
\ead{frere@ulb.ac.be}

\begin{abstract}
In these notes, we review the main results of our approach to fermion
masses. The marge mass ratios between fermions, confronted with a unique
breaking mechanism leading to vector bosons masses, led us to consider the
possibility that they result from the overlap of fermion wave functions.
Such overlaps vary indeed very strongly if the observed fermion families
in 4 dimensions originate in a single family in 6 dimensions, through
localized wave functions. This framework leads in a natural way to large
mass ratios and small mixing angles between quarks. What came as a
surprise is that if we impose that neutrinos behave as 2-component
(``Majorana'') particles in 4D, a completely different situation is
obtained for them. Instead of diagonal mass matrices, anti-diagonal ones
emerge and lead to a generic prediction of combined inverted hierarchy,
large mixing angles in the leptonic sector, and a suppression of
neutrinoless-double beta decay placing it at the lower limit of the
inverted hierarchy branch, a challenging situation for on-going and
planned experiments. Our approach predicted the size of the $\theta_{13}$
mixing angle before its actual measurement. Possible signals at colliders
are only briefly evoked.
\end{abstract}

\section{Introduction}
The spectrum of masses and mixing of fermions (with masses extending from 0.5 MeV to 170 GeV for the charged ones,
while the light neutrinos have masses below 1 eV) is difficult to account for, and calls, in the basic formulation of the Standard
Model, for a wide range of Yukawa couplings to the (Brout-Englert-Higgs) scalar sector. One more striking feature is
the weakness of the mixing between families of quarks, opposed to the unexpectedly strong ones in the leptonic sector.

We have studied for some time the possibility that such a spectrum originates in fact from a framework where the 3 known families
in 4  (3+1) dimensions originate from a single one in a 6-dimensional world.
This work was initally centered on the quark sector, but when extending it to the leptonic sector we realized an unexpected result,
namely, insisting on Majorana-like neutrinos in 4D implied naturally large mixings, inverted hierarchies, suppression of the neutrinoless
double beta decay to a value towards the minimum allowed in inverted hierarchy, and a measurable (but at the time yet unmeasured)
third mixing angle in the leptonic sector $\theta_{13}$ .

The following sections constitute an overview of the approach and its progress toward the current description, but for detailed
calculations and full Lagrangians of the model, we refer the reader to the original papers.

\section{From plane to sphere}
The initial formulation of our approach took place with the 2 extra dimensions considered to form a plane \cite{LT,FLT,FLT_neutrino,LN_fit}.
In this plane,
fermions, gauge and scalar bosons propagate freely, but a scalar field
$\Phi$ introduces a topological singularity, as is customary in models
with large extra dimensions (see, e.g., Ref.~~\cite{Rubakov} for a review).
Indeed, the potential of $\Phi$ is chosen such that it develops a vacuum
expectation value, and the topology assumes that it behaves like $e^{i n
\phi}$
%
%at large distance from the origin
, where $ (R, \phi)$ are the polar
coordinates in the extra-dimensions plane. The field  $\Phi$ vanishes at
the origin (to avoid a multiple determination), and a finite-energy
solution is obtained with the help of an auxiliary (axial) field $A$
coupling both to  $\Phi$ and to the matter fields involved.
The fermions coupled to  $A, \Phi$ can be shown to possess "zero modes" in
the 2 extra dimensions, meaning that they lead to massless particles in
4D \cite{JackiwRossi, Witten-SCstrings}. The number of such zero modes,
for each 6D fermion type, can be shown to be equal to $n$, the "winding
number" of the vortex solution so obtained.  There is no special reason,
except phenomenology, to set $n=3$ (although we realized later that a
winding equal to 4, for instance, would lead to an unacceptable neutrino
phenomenology), but this value allows to generate the 3 observed families
of fermions from just one family in 6D.

One major difficulty with the "plane" extra 2D is that the fermions are
localized close to the core of the vortex, (and similarly the scalars
coupled to the vortex), while no such localization occurs for the gauge
bosons. Since the effective coupling in 4D is driven by the overlap of the
wave functions, this would result in infinitesimal coupling of fermions to
gauge bosons. There are two ways out, one being to invoke some
(superconductivity-inspired, for instance~\cite{DvaliShifman}) mechanism
for localizing the gauge particles, but it proves difficult. The simplest
way is then to consider the extra 2 dimensions as spanning a sphere,
rather than the infinite plane. Although the implementation of the vortex
is slightly more thechnical, the changes in terms of phenomenological
results are minimal \cite{FLNT_sphere,FLNT_flavour,FLNT_LHC}. A major
result was established when investigating the possibility of generating
Majorana-type neutrinos in 4 dimensions in the context of our 6
dimensional approach. Not only did we show that this was possible
(although there are no Majorana particles in 6D), but it also resulted in
a link between the Majorana character and the unusual mixing properties of
the leptonic sector \cite{FLL_neutrino,Frere:2013eva}.

\section{Fermion fields and vortex profiles, electroweak symmetry breaking}
In 6 dimensions, the Lorentz group is represented on 8-component spinors, which can further (for massless particles) be split into two
inequivalent representations, according to the $\pm$ eigenvalues of a
matrix $\Gamma_7$ (6D chirality). With compactification in mind, the
resulting 4-component spinors $\Psi_{\pm}$ will later appear as 4D Dirac
spinors, composed of the more usual $L$ and $R$ chiralities.
\be
\Psi = \left(
\ba{c}
\psi_{+R} \\ \psi_{+L} \\ \psi_{-L} \\ \psi_{-R}
\ea
\right) \quad .
\ee
When coupling the 6D fermions to the winding-3 scalar field, 3 massless modes emerge, labeled $n=1,2,3$.
It should be noted that they correspond to \emph{chiral} fermions in 4D. Indeed a single 2-component spinor $\psi$
appears, although it comes in a redundant way (it is present in both the $+$ and $-$ parts of the original spinor).
We can in this way generate for each 6D fermion 3 chiral fermions in 4D, corresponding to the 3 families of particles.
More precisely, we obtain the left-handed fields:

\be
L \sim \sum_n \left(
\ba{c}
0\\ f_{3-n}(r) \ e^{i(3-n)\phi}  \psi_{Ln}(x^{\mu}) \\ f_{n-1}(r) \ e^{i(1-n)\phi}  \psi_{Ln}(x^{\mu})  \\ 0
\ea
\right) \quad ,
\ee

Similarly, we can generate right-handed fermions by adapting the coupling to the vortex field $\Phi$:

\be
R \sim \sum_n \left(
\ba{c}
\\  f_{n-1}(r) \ e^{i(1-n)\phi}  \chi_{Rn}(x^{\mu})   \\ 0  \\ 0\\f_{3-n}(r) \ e^{i(3-n)\phi}  \chi_{Rn}(x^{\mu})
\ea
\right) \quad ,
\ee

It must be noted that each of the "zero modes" possesses a very specific dependence in $\phi$, the polar angle in the remaining 2D,
but also that their radial dependance is very different. This will result, on one side ($\phi$ dependance) to very specific forms for the
mass matrix, and on the other side, to very different masses for the different fermion families.

The electroweak symmetry breaking occurs in a very similar way to the 4-dimensional case, through an explicit Standard Model Scalar
(Brout-Englert-Higgs fiels) H. The difference here is that the scalar field is bound to the vortex, and its vacuum expectation value is thus
dependant on the extra dimensions. In the present case, we chose a situation where it carries no winding number.

The field assignments are listed in Table \ref{tab:charges}, note that we included an extra scalar field $X$, which is used
to generate the (small) off-diagonal elements of the mass matrix. For details see \cite{Frere:2013eva}, note however
that instead of introducing products of fields (like $\overline{e_R} H^{\dag} X \Phi^*  L$) in \cite{Frere:2013eva}, we could choose to use
additional scalar doublets $H_i$.

\begin{table}[h!]
\centering
\begin{tabular}{|c|c|c|c|c|c|}
\hline\hline Field& Notation & $U(1)_g$ & $U(1)_Y$ & $SU(2)_w$    &
$SU(3)_c$ \\
\hline \hline vortex scalar&$\Phi$ & $+1$     & $0$      & $\mathbf{1}$
& $\mathbf{1}$\\
\hline BEH boson&$H$ & $0$      & $+1/2$   & $\mathbf{2}$ &
$\mathbf{1}$\\
\hline auxiliary scalar&$X$    & $+1$     & $0$      & $\mathbf{1}$ &
$\mathbf{1}$\\
\hline \hline quark $SU(2)_{\rm W}$ doublet&$(Q_{+},Q_{-})$ & $(3,0)$
& $1/6$ & $\mathbf{2}$ & $\mathbf{3}$\\
\hline up-type quark $SU(2)_{\rm W}$ singlet&$(U_{+},U_{-})$    &
$(0,3)$ & $2/3$ & $\mathbf{1}$ & $\mathbf{3}$\\
\hline down-type quark $SU(2)_{\rm W}$ singlet&$(D_{+},D_{-})$    &
$(0,3)$  & $-1/3$ & $\mathbf{1}$ & $\mathbf{3}$\\
\hline lepton $SU(2)_{\rm W}$ doublet&$(L_{+},L_{-})$    & $(3,0)$  &
$-1/2$   & $\mathbf{2}$ & $\mathbf{1}$\\
\hline charged lepton $SU(2)_{\rm W}$ singlet&$(E_{+},E_{-})$    &
$(0,3)$ & $-1$     & $\mathbf{1}$ & $\mathbf{1}$\\
\hline sterile neutrino singlet&$(N_{+},N_{-})$    & $(0,0)$  & $0$
& $\mathbf{1}$ & $\mathbf{1}$\\
\hline\hline
\end{tabular}
\caption{
\label{tab:charges}
Charge assignments of the fields under the gauge
groups of the SM and under $U(1)_{\rm g}$. For fermions, the two numbers
in parentheses correspond to the charges of the components with positive
and negative values of the $\Gamma_{7}$ parity, denoted everywhere by
``$+$'' and ``$-$'' indices, correspondingly.}
\end{table}

\begin{figure}
\begin{center}
\includegraphics[width=10cm]{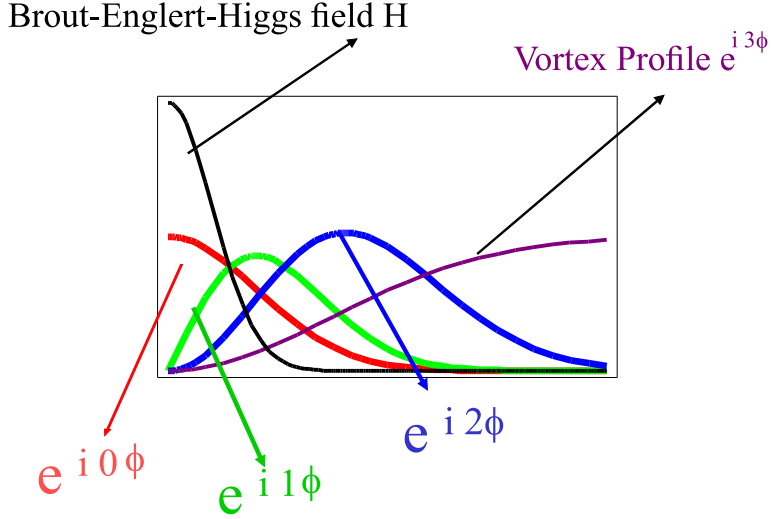}
\end{center}
\caption{\label{profiles}Profiles of the charged fermions, $H$ and vortex field $\Phi$ in the extra dimensions.}
\end{figure}

The respective profiles of the charged fermions and the scalar field $H$ are centered on the vortex,
and an approximate picture is given in Figure \ref{profiles}.

\section{Fermion masses - the good news: neutrinos are different!}
Fermion masses, as announced, result from the convolution of the fermion fields (expressed in 6 dimensions) with
the scalar boson field.
For charged fermions, the (main) couplings are of the type $e_R H^\dag L$, and the $\phi$ integration leads to
a diagonal matrix (the scalar field carries no winding number), with strongly hierarchical masses (as it has
the strongest overlap with the mass-giving $\langle H \rangle$, the fermion
with winding number = 0 gets the largest mass). The winding number $n$
acts as an effective family number $f$ with $f=3-n$.

The situation is completely different for neutrinos, if we insist that they behave in 4 dimensions as Majorana (Weyl) fields.
The masses of our "light" neutrinos result, through a see-saw type
mechanism, from coupling to the "bulk" field $N$ (the equivalent of the
right-handed neutrino in 4 dimensions). A bulk mass term $M \
\overline{N^c} N$ induces a similar contribution for the light neutrino,
namely $\overline{\psi_{\nu}^c} \ \psi_{\nu}$, where $\psi^c$ is the
charge conjugated field. As is readily seen from equation (2) this leads
to an off-diagonal mass matrix in family space:
\begin{equation}
\left(
  \begin{array}{ccc}
    0 & 0 & m\\
   0 & \mu& 0 \\
    m & 0 & 0 \\
  \end{array}
\right) ,
\end{equation}
where the dynamics leads to $\mu \ll m$, while the matrix is symmetrical by construction.

The consequences of this shape are striking. It implies a maximal mixing between neutrinos 1 and 3 (in our
notations), but also a degeneracy between the heavier 2 neutrinos.

Of course, this cannot be the whole picture, and extra contributions are needed, both in the charged fermions and
neutrino sectors, to provide the full mixing pattern (these smaller contributions are brought in an effective way
in our model by the $X$ field); yet, the pattern we have here, after diagonalisation of the neutrino matrix, leading to
\begin{equation}
\left(
  \begin{array}{ccc}
    m & 0 & 0\\
   0 & -m& 0 \\
    0 & 0 & \mu \\
  \end{array}
\right) ,
\end{equation}
points to a set of generic predictions:
\begin{itemize}
  \item inverted hierarchy of neutrinos (2 heavier, closely-spaced
neutrinos, and a lighter one);
\item large mixing angles (one comes automatically and is linked to the
Majorana nature of the 4D neutrinos);
\item suppression of the lepton number violation: equal and opposite masses $m, -m$ lead to cancelation of
  contributions to the neutrinoless-double beta decay: this is a "pseudo-Dirac" scheme, involving neutrinos from
  2 families.
\end{itemize}
This complements the already-made generic features of the quark (and to some extend charged leptons) sectors,
namely:
\begin{itemize}
  \item nearly diagonal mass matrices;
  \item strong hierarchy of charged fermions masses between the 3 families.
\end{itemize}

\section{Predictions, prospects and experimental constraints}
A full description of the masses and mixings necessitates to introduce more (off-diagonal) couplings, and therefore
leads to more arbitrariness. We constructed somewhat minimal models (with a limited number of off-diagonal terms)
and found reasonable mass spectra and mixings.

One definite and now verified  prediction was for a non-zero $U_{e3}$ of
order  $\sin(\theta_{13}) \simeq 0.13$, before its measurement by the Daya
Bay experiment \cite{An:2012eh}.

In terms of checking the Majorana nature of the neutrinos, neutrinoless double beta decay are the only accessible test:
the "pseudo-Dirac" partial cancelation puts our prediction at the lower bound of the "inverted hierarchy case", with
an effective mass $m_{\beta  \beta} \simeq 0.013$ eV
which makes it reachable only for a later stage of experiments \cite{Gerda}.

In terms of accelerator physics, the prospects are less encouraging, at least in the basic formulation of our approach.
Indeed, the extra dimension scale $R$ is associated to the mass of extra gauge bosons in 4 dimensions (Kaluza-Klein excitations).
In our case, this involves $Z'_0$, $Z'_{\pm 1}$ (and similar states for the
photon and gluons). The $Z'_{\pm 1}$ cary on unit of winding, which means
that they are flavour changing, but conserve (in the first, diagonal
matrix, approximation) family number.

A clean test is given by $K^0_L \rightarrow \overline{e} \mu $, which
provides a limit
\begin{equation}
\kappa  / M^2_{Z'} < (100 \mbox{TeV})^{-2}.
\end{equation}
Here, $\kappa$ measures the overlap between the wave functions of the
fermions pertaining to different families. In our present analysis (we are
working on alternate cases), its value does not allow to bring the $1/R$
scale much below 100 TeV, but it is conceivable that other geometries
would allow it. In that case $\overline{\mu} e$ (in excess by one order of
magnitude over $\mu \overline{e}$) would provide a particularly striking
signal \cite{FLNT_flavour,FLNT_LHC}.

\ack
This work is funded in part by IISN and by Belgian Science Policy (IAP
"Fundamental Interactions").
The work of ML and ST (elaboration of the model of the origin and hierarchy of neutrino masses and mixings in the context of new experimental data) is supported by the Russian Science Foundation, grant 14-22-00161.

\smallskip

\end{document}